\documentclass[preprint,12pt]{elsarticle}

\usepackage{amssymb}
\usepackage{amsmath}

\journal{NIM A Proceedings of RICAP 2011}

\begin{document}

\begin{frontmatter}



\title{Dark matter search with the ANTARES neutrino telescope}


\author{Juan de Dios Zornoza\\
{\small on behalf of the ANTARES collaboration}}

\address{IFIC, Ed. Institutos de Investigación, AC 20085, E-46071, Valencia, Spain}

\begin{abstract}
The ANTARES neutrino telescope was completed in 2008 with the installation of its twelfth line. Its scientific scope is very broad, but the two main goals are the observation of astrophysical sources and the indirect detection of dark matter. The latter is possible through neutrinos produced after the annihilation of WIMPs, which would accumulate in sources like the Sun, the Earth or the Galactic Centre. The neutralino, which arises in Supersymmetry models, is one of the most popular WIMP candidates. KK particles, which appear in Universal Extra Dimension models, are another one. Though in most models these annihilations would not directly produce neutrinos, they are expected from the decay of secondary particles. An important advantage of neutrino telescopes with respect to other indirect searches (like gamma rays or cosmic rays) is that a potential signal (for instance from the Sun) would be very clean, since no other astrophysical explanations could mimic it (like pulsars for the case of the positron excess seen by PAMELA). Moreover, the Galactic Centre is accessible for ANTARES, being in the Northern Hemisphere. In this talk I will present the results of the ANTARES telescope for dark matter searches, which include neutralino and KK particles.
\end{abstract}


\end{frontmatter}

\section{Introduction}
\label{intro}

The nature of dark matter is one of the hottest mysteries in physics nowadays. Different observations, like the data obtained by the WMAP satellite~\cite{wmap} combined with the studies of highly red-shifted Ia supernovae~\cite{snia}, conclude that the Universe is made of a 73\% of dark energy and 27\% of matter. Moreover, several observational facts point out that most of the matter in the Universe is not luminous. For instance, galaxies rotate as if they contained much more matter that what is actually seen emitting radiation. Different experiments show both that the component of luminous dark matter is $\Omega h^{2} \sim 0.01$ and that the total contribution from matter is 0.0945 $< \Omega h^{2} < $0.1287. The models which explain the Big Bang nucleosysthesis put a limit in the barion density in the Universe ($\Omega h^2 \sim 0.02$), taking into account measurements of the abundances of helium, deuterium and lithium. Therefore, a great contribution of non-barionic dark matter is needed. One natural candidate are neutrinos, but since they travel at relativistic velocities, they could not contribute to the formation of the matter structures as seen in the Universe today. Therefore, the candidate for explaining the dark matter must be outside the Standard Model. The most accepted models assume the existence of Weakly Interacting Massive Particles (WIMPs). The required properties of such particles (stable, massive and interacting with a cross-section similar to that of the weak interactions) can be produced in several models. Here we are considering two kind of models: Supersymmetry (SUSY) and Universal Extra Dimensions (UED). In both cases, the stability of the dark matter candidates is guaranteed by a symmetry conservation (R-parity in the case SUSY and KK-parity in the case of UED). More specifically, we will use the subset of models in which some extra assumptions have been made: CMSSM and mUED, for which the dark matter candidates are the neutralino and the lightest Kaluza-Klein particle (see~\cite{bertone} for a complete review). In these models, the WIMPs would accumulate in massive objects like the Sun, the Earth and the Galactic Centre after losing energy through elastic scattering and become gravitionally trapped. If they are Majorana particles, they would self-annihilate and produce directly or indirectly neutrinos and other particles. Therefore, neutrino telescopes can observe these objects and look for an excess of neutrinos over the expected background. The case of the Sun is particularly interesting, since a potential signal would be very clean, in the sense that it could hardly be explained by other astrophysical phenomena (the neutrinos produced in the nuclear reactions are of low energy and the background due to neutrinos produced by the interaction of cosmic rays on the Sun corona is negligible). In this paper we will present the analysis made in the ANTARES neutrino telescope to search for a neutrino signal in the Sun direction, using a binned method (i.e. counting events within a given angular cone around the Sun position) and data of 2007-2008.

\section{The ANTARES neutrino telescope}
\label{antares}

The installation ANTARES neutrino detector~\cite{antares} was completed in 2008. It is located at a depth of 2475~m in the Mediterranean Sea, 42 km from Toulon, in the French coast (42$^{\circ}$48 N, 6$^{\circ}$10 E). This detector consist of a tridimensional array of 885 10-inch photomultipliers (PMTs) distributed along twelve detection lines. Each PMT is housed in a pressure-resistant glass sphere, the so-called Optical Module (OM). The OMs are grouped in 25 triplets (or storeys) per line except for one of the lines where some devices for accoustic detection are installed and therefore contains only 20 storeys. The PMTs are pointing 45$^{\circ}$ downwards in order to increase the performance to detect up-going events. The vertical distance between consecutive storeys is 14.5~m. The lines are anchored at the sea floor and kept vertical by means of buoys at the upper part of the line. The separation between lines is 60-70~m. The detector is also complemented with a line with additional instrumentation for environmental measurements. ANTARES is presently the largest neutrino telescope in the northern hemisphere.

The operation principle is as follows: when a high energy neutrino interacts via charged current inside the detector or close to it, it produces a relativistic muon which, in the water, induces Cherenkov light observable by the PMTs. The information of the time~\cite{timing}, position and amplitude of the photon signals in the PMTs is used in order to reconstruct the muon track and therefore the direction of the original neutrino. Other signatures are also possible, like cascades produced both by electron and tau neutrinos in CC interactions and in NC interactions of all neutrino flavours. The muon track is reconstructed with the algorithm BBFit~\cite{bbfit}, which provides an angular resolution of about two degrees at energies of tens of GeVs for the selected tracks. The variable $\chi^2_t$ is related with the quality of the reconstructed tracks and will be used in the cut optimization, as explain in Section~\ref{cuts}.

\section{Data and Monte Carlo simulation}
\label{simulation}

\begin{figure*}
\includegraphics[width=0.5\linewidth]{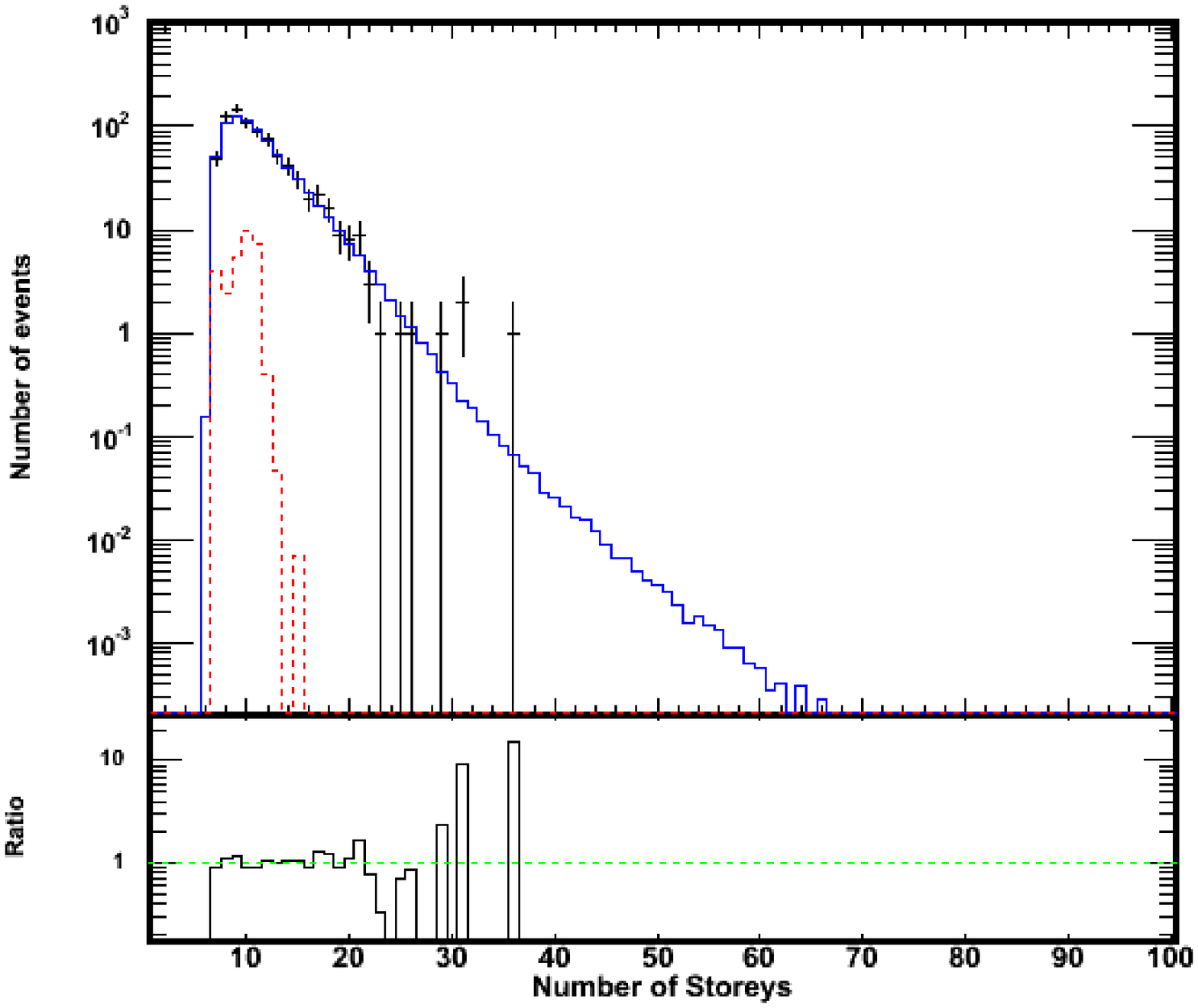}  
\includegraphics[width=0.5\linewidth]{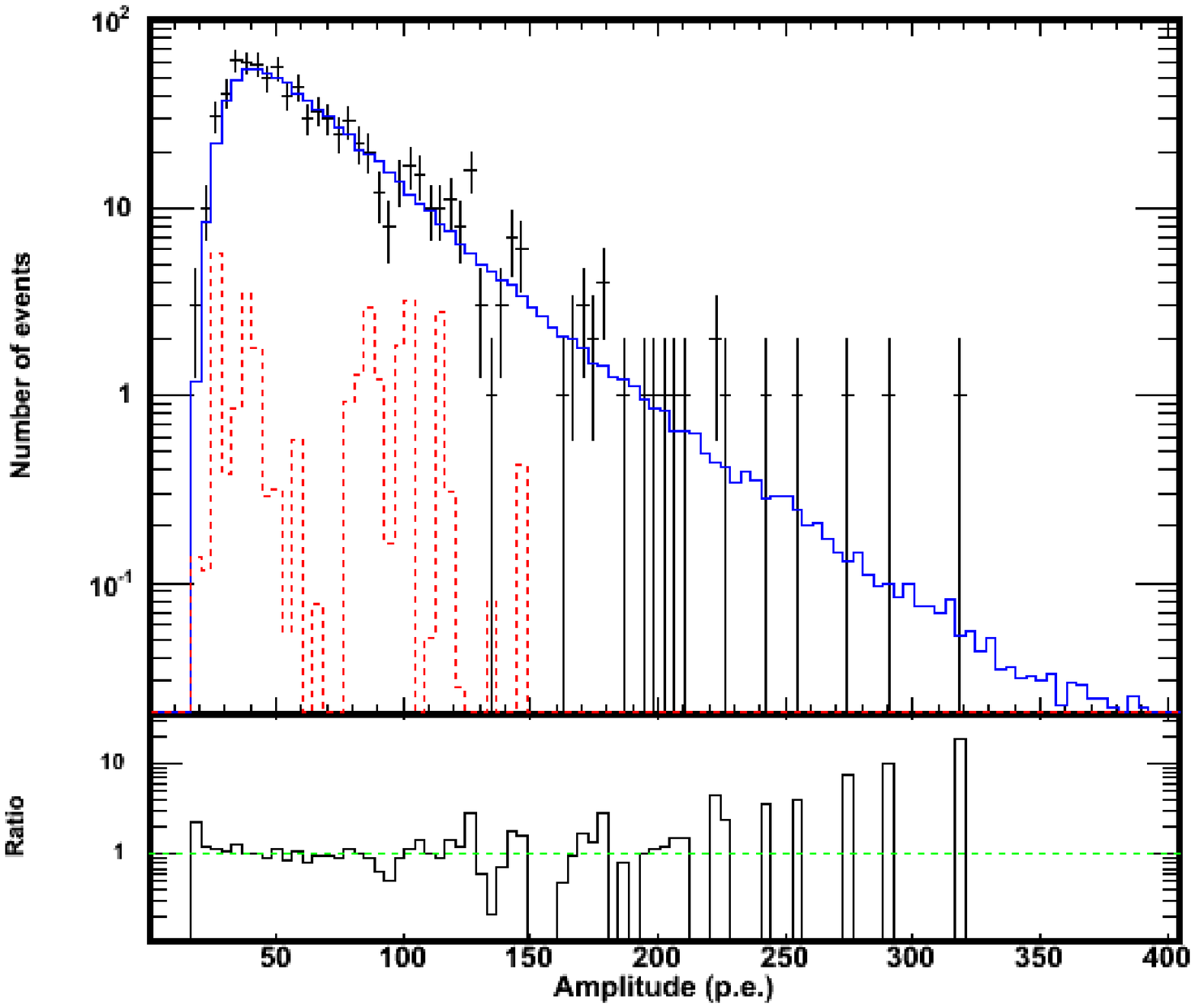}  \\
\includegraphics[width=0.5\linewidth]{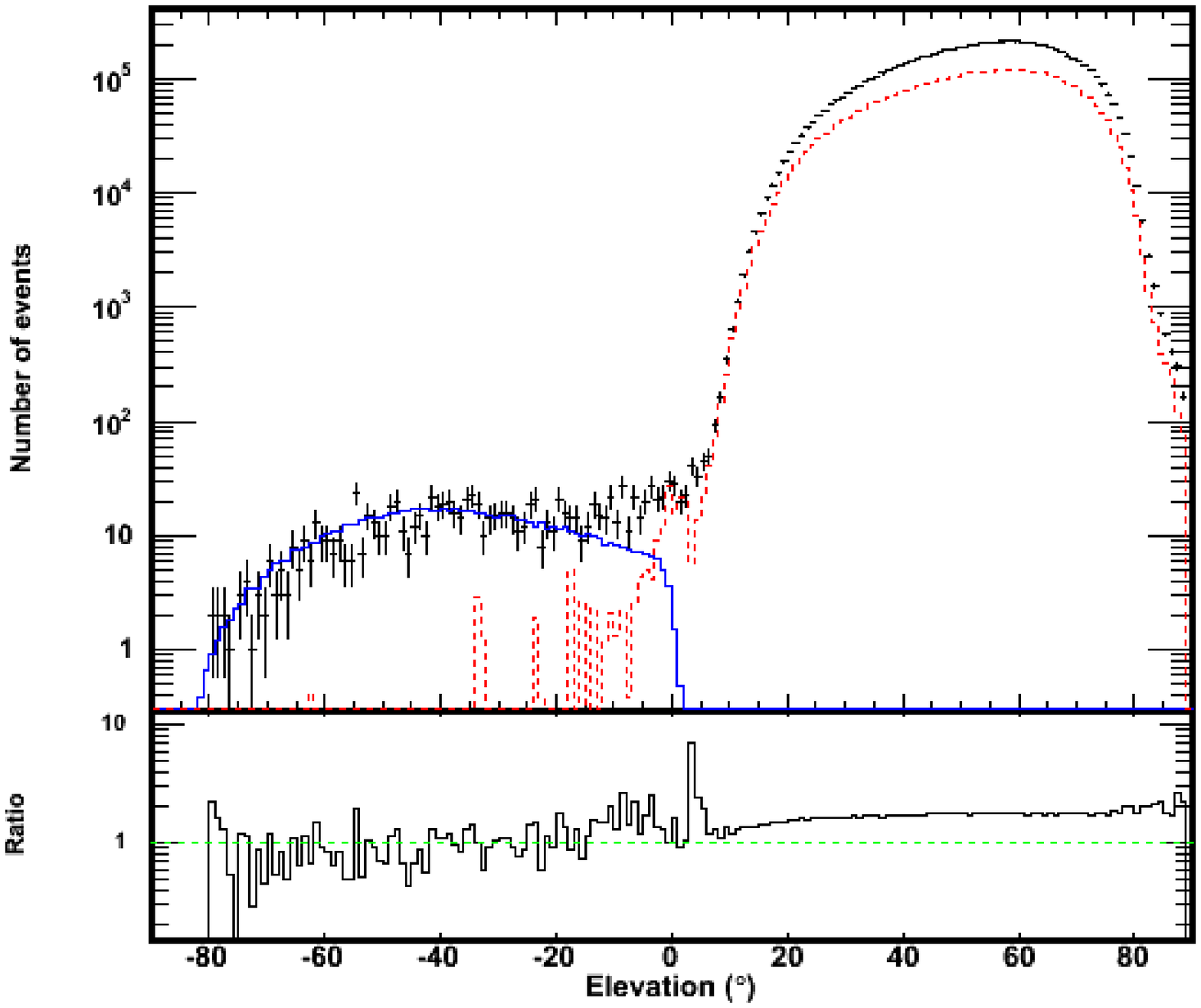} 
\includegraphics[width=0.5\linewidth]{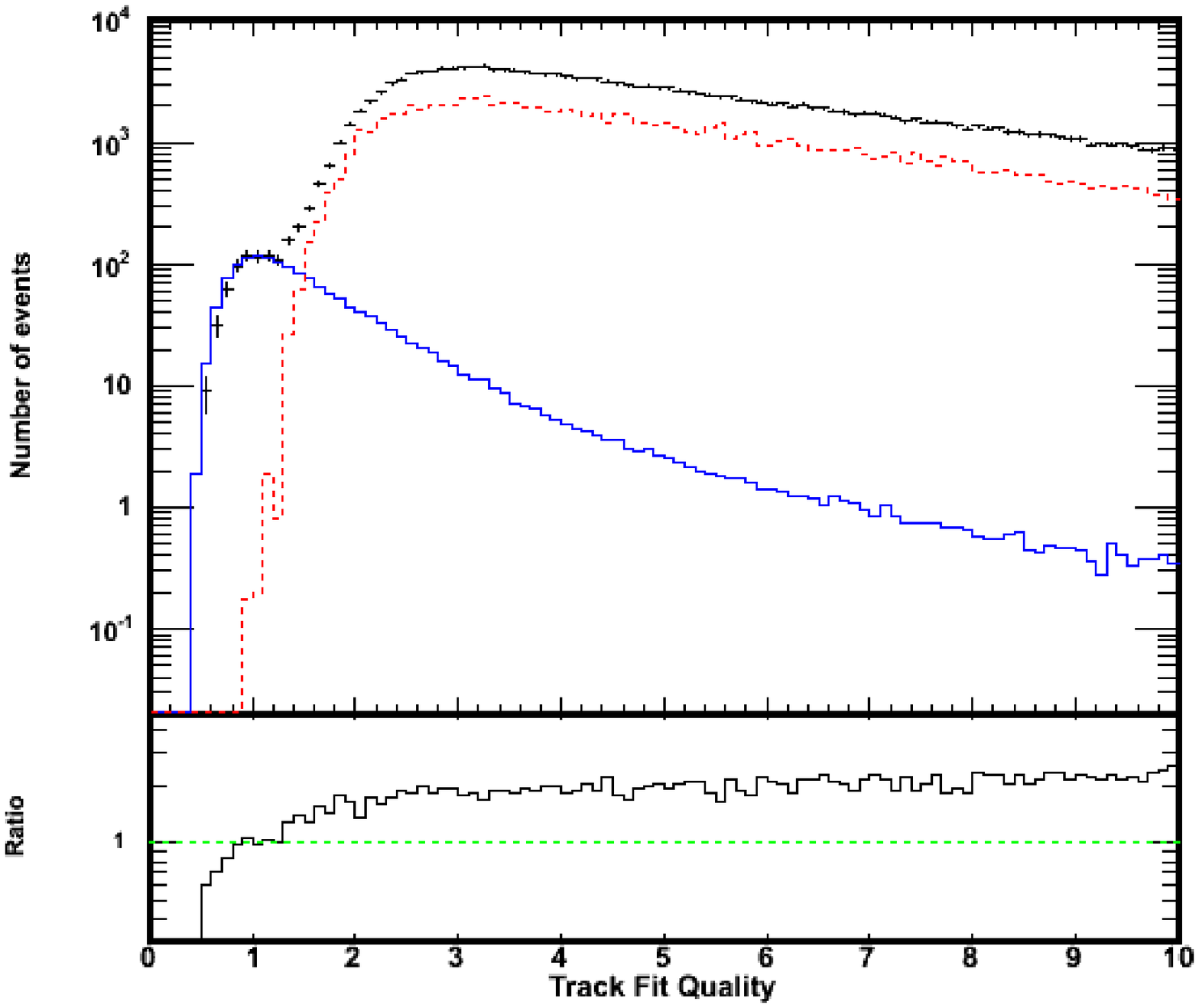} \\
\caption{Comparison between data and simulation: number of storeys (top left),  amplitude (top right), elevation (bottom left) and track fit quality parameter $\chi^2_t$ (bottom right). Except in the last case, $\chi^2_t < 1.4$ cut has been applied. The distributions show the simulated atmospheric muons (red), the simulatd atmospheric upgoing-neutrinos (blue), and the reconstructed data (black). In each case, the ratio of data over simulation is shown below the corresponding plot.}
\label{mcdata}
\end{figure*}

The data used for this analysis correspond to the period from 2007 to 2008. During almost all of 2007, only 5 lines of the detector where installed (185.5 active days). During 2008, the configuration of the detector consisted in 10, 9 or 12 lines (189.8 active days in total). Figure~\ref{mcdata} shows a comparison between data and simulation for several relevant distributions. There are two sources of background in ANTARES. On the one hand, the huge flux of down-going atmospheric muons, produced in the interaction of cosmic rays in the atmosphere. Although a large part of this background is reduced due to the fact that ANTARES is located deep in the sea, there is still a large amount of down-going muons arriving at the detector. In order to further reduce them, only events reconstructed as up-going are considered. Even with such selection, it is possible that some down-going events are reconstructed as up-going, so a selection based in the quality of the reconstructed track is also made. The other kind of background is due to the atmospheric neutrinos produced in the atmosphere by cosmic rays. These neutrinos can traverse the Earth, so they are detected as upgoing tracks. This is an irreducible background. The way to deal with it is based on the fact that it is a diffuse flux, while the signal is expected to peak in the direction of the Sun position. For the estimation of the background we have used scrambled data. This allows to reduce the effect of systematic uncertainties (efficency of the detector, assumed flux, etc.)

The other ingredient of the analysis is the simulation of the signal. This is done in three steps. First we use the software package WimpSim~\cite{wimpsim} which computes the differential amount of neutrinos per energy bin which are created in the Sun core and arrive at the Earth surface. This package allows the generation of all possible channels, as quarks/anti-quarks, leptons/anti-leptons, Higgs doublets and the possible direct production channels in neutrinos of the three flavours. Therefore, it allows to reproduce any given model, provided the corresponding branching ratios are known. The main physical processes are taken into account, like the interactions in the Sun medium, regeneration of tau leptons, oscillations in the propagation, etc. The second step then is to weight the fluxes to simulate the model under analysis. As mentioned before, we have considered two cases: CMSSM and mUED. The main channels in a neutrino telescope are $W^+W^-$, $b\bar{b}$ and $\tau\bar{\tau}$ for CMSSM and $c\bar{c}$, $b\bar{b}$, $\tau\bar{\tau}$, $t\bar{t}$ and $\nu\bar{\nu}$ for mUED. Finally, a third step consists in calculating the number of events expected in the detector, through a simulation of the detector response to the neutrino flux (which in turn depends on the energy, track direction and selection cuts applied). For this, the standard software used in the ANTARES collaboration is used. It is also worth to mention that even if the angular diameter of the Sun is about half degree, the signal is expected to be produced in the Sun core, so we can consider the Sun as a point source. Moreover, equilibrium between capture and annihilation in the Sun is assumed.

\section{Cut optimization}
\label{cuts}

\begin{figure}
\includegraphics[width=\linewidth]{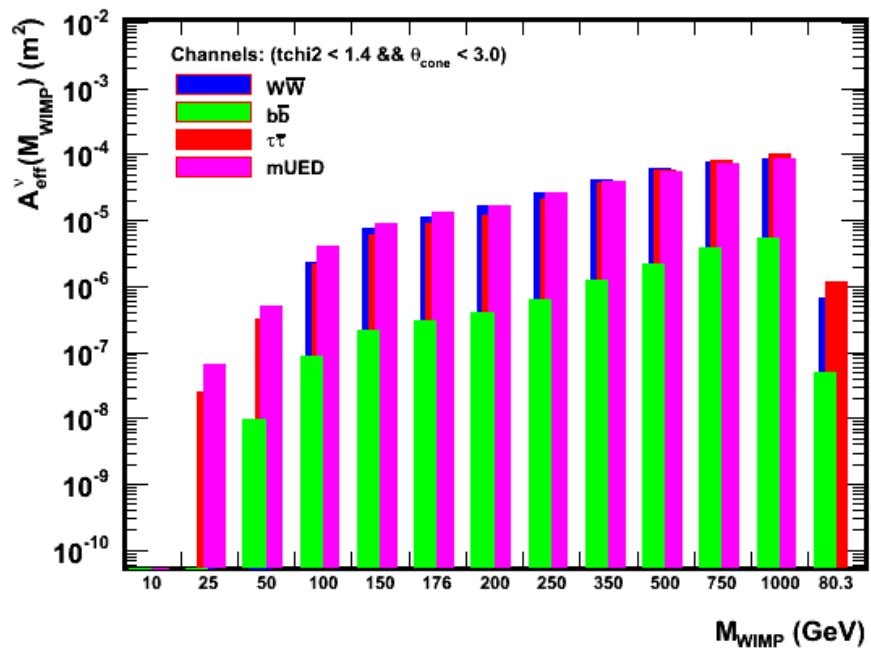}
\caption{Normalized effective area for different WIMP masses. The particular case of $M_{WIMP}=80.3$ GeV, corresponding to the beginning of the $W^+W^-$ contribution, is also shown.}
\label{aeff_channels}
\end{figure}

In this analysis we follow a blinding policy in order to avoid biases in the event selection. This means that the values of the cuts are chosen before looking at region where the signal is expected. The optimization of the cuts is done using the Model Rejection Factor (MRF) technique~\cite{hill}. It consists of finding the set of cuts which provide, in average, the best flux upper limit. The two parameters used for the optimization are the quality cut of the track fit reconstruction algorithm, $\chi^2_t$, and the half-cone angle around the Sun, $\Psi$. This limit can be expressed as 

\begin{equation}
\bar{\phi}_{\nu}^{90\%} = \frac{\bar{\mu}^{90\%}}{A_{eff}(M_{wimp}) \times T_{eff}}, 
\label{phinuaulimiteqn}
\end{equation}

\noindent where $\bar{\mu}^{90\%}$ is the average limit in the number of events, $A_{eff}(M_{WIMP})$ is the effective area and $T_{eff}$ is active time of data taking. The effective area is defined as the area so that a detector which would be 100\% efficient would produce the same number of detected events. We have included in it also the visibility of the Sun, which is the fraction of time where the Sun is visible, and the bin efficiency, which is the fraction of signal kept by the angular cut. Figure~\ref{aeff_channels} shows the effective area for $\chi^2_t < 1.4$ and $\Psi < 3^{\circ}$

Depending on the considered channel and WIMP mass, different set of these parameters optimize the average upper limit (Table~\ref{opcut}).

\begin{table*}
\begin{center}
\begin{tabular}{|c|c|c|c|c|c|c|c|c|}
\hline
{} & \multicolumn{2}{c|}{Channel $W^{+}W^{-}$} & \multicolumn{2}{c|}{Channel $b\bar{b}$} & \multicolumn{2}{c|}{Channel $\tau\bar{\tau}$} & \multicolumn{2}{c|}{Channel mUED} \\
\hline
{$M_{wimp}$ (GeV)} & {$\chi^2_t$} & {$\psi$ ($^{\circ}$)} & {$\chi^2_t$} & {$\psi$ ($^{\circ}$)} & {$\chi^2_t$} & {$\psi$ ($^{\circ}$)} & {$\chi^2_t$} & {$\psi$ ($^{\circ}$)} \\
\hline
{50} & {-} & {-} & {1.1} & {3.5} & {1.3} & {5.8} & {1.3} & {5.6} \\
{80.3} & {1.6} & {6.0} & {1.3} & {5.6} & {1.3} & {5.6} & {-} & {-} \\
{100} &  {1.4} & {5.0} & {1.3} & {5.6} & {1.3} & {5.6} & {1.3} & {5.6} \\
{150} &  {1.3} & {5.6} & {1.3} & {5.6} & {1.3} & {5.6} & {1.4} & {4.5} \\
{176} &  {1.4} & {4.5} & {1.3} & {5.6} & {1.4} & {4.5} & {1.4} & {3.5} \\
{200} &  {1.4} & {3.9} & {1.3} & {5.6} & {1.4} & {4.5} & {1.3} & {4.5} \\
{250} &  {1.4} & {3.9} & {1.3} & {5.6} & {1.4} & {4.5} & {1.3} & {4.5} \\
{350} &  {1.4} & {3.6} & {1.3} & {5.6} & {1.4} & {3.9} & {1.5} & {3.9} \\
{500} &  {1.4} & {3.6} & {1.3} & {4.5} & {1.4} & {3.6} & {1.4} & {3.9} \\
{750} &  {1.4} & {3.6} & {1.3} & {4.5} & {1.4} & {3.6} & {1.4} & {3.6} \\
{1000} &  {1.5} & {3} & {1.4} & {4.5} & {1.4} & {3.6} & {1.4} & {3.0} \\
\hline
\end{tabular}
\caption{Values of the cuts in the quality reconstruction parameter $\chi^2_t$ and in the half-angle search cone $\Psi$ which optimize the sensitivity for different WIMP masses and channels.} 
\label{opcut}
\end{center}
\end{table*}

\section{Results}
\label{results}

\begin{figure}
\includegraphics[width=\linewidth]{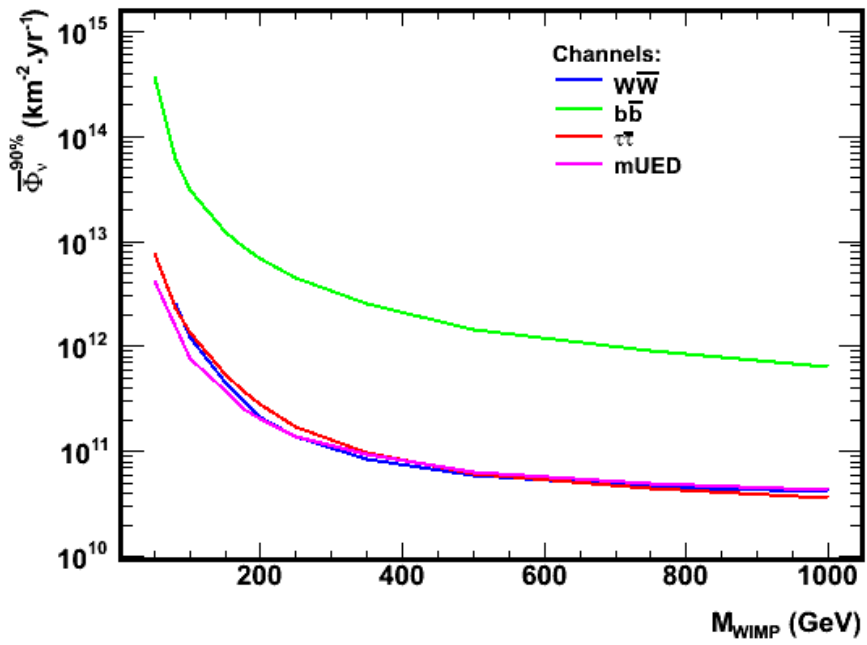}
\caption{Sensitivity in the neutrino flux as a function of the WIMP mass for different cases: $W^+W^-$ (blue), $b\bar{b}$ (green), $\tau\bar{\tau}$ (red) and mUED (magenta). (Preliminary)}
\label{fluxlimit}
\end{figure}

Once the selection cuts have been optimized, we can estimate the sensitivity for the different models, by using the equation \ref{phinuaulimiteqn}. The corresponding average upper limits on the neutrino flux are shown in figure~\ref{fluxlimit}. This figure shows the average upper limit in the neutrino flux as a function of the WIMP mass. Given the wide spread in the branching ratios in the CMSSM models, we present separatedly the main contributing channels. For the mUED models, this spread is much lower, so we can merge the channels for a particular realization of this model since this should be a good representation of all the choices for the parameters of the model. As it can be seen in the plot, for CMSSM, the channels $W^+W^-$ and $\tau\bar{\tau}$ produce the best limits, since the amount of neutrinos expected in these channels is higher. For mUED, the most important contribution to the signal comes from the $\tau\bar{\tau}$ channel, so we obtain a sensitivity very similar to the one corresponding to only that channel.

Finally, the obtained limits in the neutrino flux can be used to constrain the predictions of the models on the spin-dependent cross section of the interactions of the neutralinos or KK particles with the protons. The differential neutrino flux is related with the self-annihilation rate $\Gamma$ as 

\begin{equation} 
\frac{d\phi_{\nu}}{dE_{\nu}} = \frac{\Gamma}{4\pi d^{2}} \, \frac{dN_{\nu}}{dE_{\nu}}, 
\end{equation}

\noindent where $d$ is the distance from the Sun to the Earth and $dN_{\nu}/dE_{\nu}$ is the differential number of neutrino events for each channel. On the other hand, the self-annihilation rate can be related with the capture rate as

\begin{equation}
\Gamma \simeq \frac{C_{\otimes}}{2},
\label{gammaexp}
\end{equation}

\noindent where it is assumed that the equilibrium between capture and annihilation has been reached in the Sun. The capture rate in turn is related to the spin-dependent scattering cross-section between a WIMP and a proton:

\begin{multline}
C_{\otimes} \simeq 3.35 \times 10^{18} s^{-1} \times \left(\frac{\rho_{local}}{0.3 \, GeV \cdot cm^{-3}}\right) \times \\
\times \left(\frac{270 \, km\cdot s^{-1}}{v_{local}}\right) 
\times \left(\frac{\sigma_{H,SD}}{10^{-6} \, pb}\right) \times \left(\frac{TeV}{ M_{WIMP}}\right)^{2}, 
\label{captureexp}
\end{multline}


\noindent where $\rho_{local} = 0.3$ GeV$\cdot$cm$^{-3}$ is the local density of WIMPs assuming a NFW profile of dark matter in the galactic halo, $v_{local} = 270$ km$\cdot$s$^{-1}$ is the local mean velocity of WIMPs assuming a Maxwell-Boltzmann velocity distribution, $M_{WIMP}$ is the mass of the considered dark matter particle and $\sigma_{H,SD}$ is the spin-dependent scattering cross-section between a WIMP and a proton.

\begin{figure}
\includegraphics[width=0.8\linewidth]{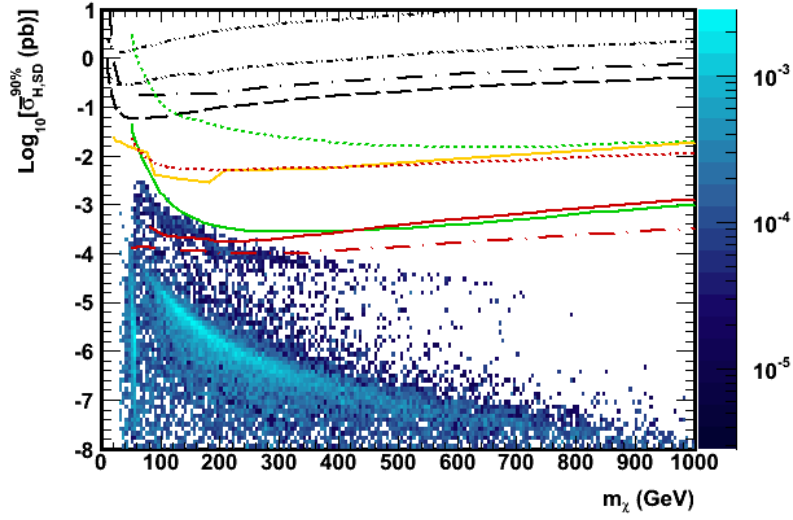}  \\
\includegraphics[width=0.8\linewidth]{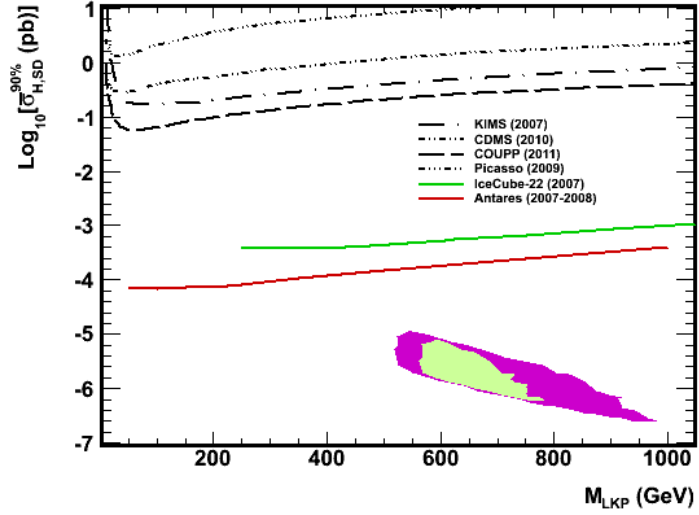}
\caption{Spin-dependent cross-section $\bar{\sigma}^{90\%}_{H,SD}$ as a function of the WIMP mass. The upper plot corresponds to the CMSSM model, with a comparison with several experimental limits: KIMS 2007 (dash-dot black line), CDMS 2010 (dash-dot-dot black line), COUPP 2011 (dashed black line), Picasso 2009 (dash-dot-dot-dot black line), IceCube-22 (dashed green line for the $b\bar{b}$ channel and solid green line for the $W^+W^-$ channel), and SuperKamiokande 1996-2001 (yellow solid line). The ANTARES sensitivity for 2007-2008 for the $b\bar{b}$, $W^+W_-$ and $\tau\bar{\tau}$ channels are shown in dashed red line solid red line and dash-dot red line, respectively. The bottom plot corresponds to the mUED model, with IceCube-22 limit in green solid line and ANTARES sensitivity in red solid line. (Preliminary)}
\label{final}
\end{figure}

In addition to this, we also use the package SuperBayes~\cite{superbayes} to scan the parameter space of the CMSSM and mUED, as shown in figure~\ref{final}. Although ANTARES is not competive with direct search experiments for spin-independent cross-sections, the contrary happens for the spin-dependent contribution, as can be seen when comparing these limits with those of other direct search experiments.

\section{Conclusions}
\label{conclusions}

The construction of the ANTARES neutrino telescope finished in 2008. In the analyis presented in this paper, we have used data taken in 2007 (when 5 lines were operative) and 2008 (with 10, 9 or 12 lines) in order to obtain the sensitivity of the detector in searching for dark matter in the Sun. In particular, we have considered two models: CMSSM and mUED. The method is based on a binned search looking for an excess of events with respect to the expected background. The average upper limits in the neutrino flux are also transformed into a sensitivity to the spin-dependent scattering cross-section of the WIMPs interaction with the protons in the Sun, which are compared with the parameter space allowed for these models. The results are competive with direct search experiments.

\section*{Acknowledgements}
\label{acknow}
The author acknowledges the support of the Spanish MICINN’s
Consolider-Ingenio 2010 Programme under grant MultiDark CSD2009-00064.








\end{document}